\documentclass[preprint,3p,12pt]{elsarticle}

\usepackage{amsmath,amssymb}
\usepackage{mathtools}
\usepackage{bm}
\usepackage{graphicx}
\usepackage{braket}	
\usepackage{dcolumn}
\usepackage{comment}
\usepackage{hyperref}

\newcommand{\be}{\begin{equation}}
\newcommand{\ee}{\end{equation}}

\newcommand{\lt}{\left}
\newcommand{\rt}{\right}
\newcommand{\del}{\partial}

\newcommand{\non}{\nonumber \\}

\newcommand{\fn}{\footnote}

\newcommand{\MSb}{\overline{\rm MS}}

\newcommand{\LMS}{\Lambda_{\overline{\rm MS}}}

\journal{Physics Letters B}

\begin{document}

\begin{frontmatter}

\begin{flushright}
$~$
\vspace*{-10mm}
{TU--1144, KEK--TH--2394}
\end{flushright}
\vspace*{10mm}

\title{Determination of $|V_{cb}|$ using N$^3$LO perturbative corrections to
$\Gamma(B \to X_c \ell \nu)$ and 1S masses}

\author{Yuuki Hayashi$^a$}
\author{Yukinari Sumino$^a$}
\author{Hiromasa Takaura$^b$}
\ead{htakaura@post.kek.jp}
\address{
$^a$Department of Physics, Tohoku University,
Sendai, 980--8578 Japan
\\
$^b$Theory Center, KEK, Tsukuba, Ibaraki 305-0801, Japan
}%

\begin{abstract}
We determine $|V_{cb}|$
using the third-order perturbative series for the inclusive semileptonic $B$ decay 
width and for the masses of the bottomonium 1S states.
We use the masses of $\eta_b(1S)$ and $\Upsilon(1S)$ 
as short-distance masses
and point out that there is a sizable difference of $|V_{cb}|$ between the two 1S mass schemes.
This is the dominant error of our determination and stems from 
insufficiency to describe theoretically
the observed mass splitting of the bottomonium 1S states.
We also study the significance of the $\mathcal{O}(\LMS^2)$ non-perturbative effects in HQET
with respect to the current perturbative accuracy.
Our result $|V_{cb}|=0.0425 (11)$ is consistent with the PDG value determined
from the inclusive decays and has a slightly larger error.
\end{abstract}

\end{frontmatter}

\section{Introduction}
The Cabibbo-Kobayashi-Maskawa (CKM) matrix elements are 
the fundamental Standard Model (SM) parameters
describing the mixing of quark flavors. Their precise determination is important to
test validity of the SM and to advance our understanding of the flavor structure.  
In particular, precise determination of the CKM matrix element $|V_{cb}|$
is desired to reduce the error of the unitarity triangle and 
to improve accuracy of SM predictions for flavor physics such as
the indirect CP violation (parametrized by $\epsilon_K$) \cite{Brod:2019rzc}, 
which can be sensitive to new physics contributions.
A long-standing issue also needs to be settled regarding the 
discrepancy between determinations of $|V_{cb}|$ using exclusive $B$ decays and those 
using inclusive decays $B \to X_c \ell \nu$.

Determinations of $|V_{cb}|$ from the inclusive semileptonic decays are based on the operator product expansion (OPE) in the
heavy quark effective theory (HQET). The decay width is given in the expansion in $\LMS/m_b$ by
\be
\Gamma=\frac{G_F^2 |V_{cb}|^2}{192 \pi^3} A_{\rm EW} m_b^5 \lt[C^{\Gamma}_{\bar{Q}Q}(m_c/m_b,\alpha_s)+C^{\Gamma}_{\rm kin} \frac{\mu_{\pi}^2}{m_b^2}+C^{\Gamma}_{\rm cm} \frac{\mu_G^2}{m_b^2}
+\mathcal{O}((\LMS/m_b)^3)\rt] . \label{DecayWidth}
\ee
The Wilson coefficient $C^{\Gamma}_{\bar{Q}Q}$ has been calculated to the $\mathcal{O}(\alpha_s^3)$ order 
\cite{Luke:1994yc,Trott:2004xc,Aquila:2005hq,Pak:2008qt,Pak:2008cp,Melnikov:2008qs,Dowling:2008mc},
where the highest order calculation has been done recently \cite{Fael:2020tow}.
The Wilson coefficients $C^{\Gamma}_{\rm kin}$ and $C^{\Gamma}_{\rm cm}$ have been calculated to the $\mathcal{O}(\alpha_s)$ order
\cite{Chay:1990da,Blok:1993va,Bigi:1993fe,Manohar:1993qn,Becher:2007tk,Alberti:2013kxa,Mannel:2014xza}.
$\mu_{\pi}^2$ and $\mu_G^2$ represent the 
non-perturbative matrix elements in HQET of $\mathcal{O}(\LMS^2)$,\fn{
In this paper we define $\mu_G^2$ with the reversed sign compared to  ref.~\cite{Hayashi:2021vdq}.} defined by
\be
\mu_{\pi}^2=\frac{\langle B(p) |\bar{b}_v D_{\perp}^2 b_v|B(p) \rangle}{2 m_B}, \quad
\mu_G^2=-\frac{\langle B(p) |\bar{b}_v \frac{g_s}{2} \sigma_{\mu \nu} G^{\mu \nu}  b_v|B(p) \rangle}{2 m_B} ,
\ee
where $| B(p) \rangle$ is the $B$ meson state (defined not in the infinitely heavy quark limit but for the finite quark mass) 
and its normalization 
is given by $\langle B(p) | B(p') \rangle=2 p_0 (2\pi)^3\delta^3(\vec{p}-\vec{p}')$; $m_B$ 
denotes the $B$ meson mass;
$b_v$ corresponds to a heavy quark field with four-velocity $v^{\mu}$ and is the dynamical variable in HQET; 
$D_{\perp}^{\mu}=D^{\mu}-v^{\mu} (v \cdot D)$ with $D^{\mu}=\del^{\mu}+i g_s A^{\mu}$.
$m_b$ and $m_c$ denote the pole mass of the bottom and the charm quarks, respectively.
The pole masses, $m_b$ and $m_c$, and the perturbative series for $C_{\bar{Q}Q}^{\Gamma}$, respectively,
have the $u=1/2$ renormalon uncertainties. 
It is known, however, that the $u=1/2$ renormalon uncertainty vanishes in the perturbative prediction for $\Gamma$,
that is $m_b^5 C_{\bar{Q}Q}(m_c/m_b,\alpha_s)$,
once the pole masses are rewritten in terms of short-distance masses \cite{Ball:1995wa}.
In this case, the leading renormalon in the perturbative prediction
is the $u=1$ renormalon or the $\mathcal{O}(\LMS^2/m_b^2)$ renormalon uncertainty,
which is to be canceled against the uncertainty of $\mu_{\pi}^2$.
Although the $\MSb$ mass $\overline{m}$ is a popular short-distance mass,
one usually avoids using $\overline{m}_b$ to rewrite $m_b$ because
(i) it is known that the perturbative relation
between the pole and the $\MSb$ masses exhibits slow convergence even after the $u=1/2$ renormalon is subtracted,
and (ii) the bad convergence becomes more serious as the power of $m_b$ gets larger;
note that we have $m_b^5$ in eq.~\eqref{DecayWidth}.
For this reason conventionally other short-distance masses are used
for the bottom quark.\fn{
This shortcoming in using the $\MSb$ mass, however, is expected to
be compensated if
the perturbative series is known to sufficiently high orders.}

The kinetic mass is a favored short-distance mass in the context of $|V_{cb}|$ determination \cite{Bigi:1994ga,Bigi:1996si,Alberti:2014yda}.
See ref.~\cite{Bordone:2021oof} for the latest determination. 
In the kinetic mass scheme, not only the $u=1/2$ renormalon but also the $u=1$ renormalon
is subtracted. This is done by an introduction of a factorization scheme.
In this scheme, the kinetic mass $m_b^{\rm kin}$ has the factorization scale dependence,
and so do general non-perturbative matrix elements.
In ref.~\cite{Bordone:2021oof}, the non-perturbative effects up to mass dimension three 
(or the operators of mass dimension up to five) are included in the OPE prediction.
By using the moments of the lepton energy distribution and those of the hadronic invariant mass distribution,
several parameters such as $|V_{cb}|$, the bottom quark mass, and the non-perturbative effects, 
are simultaneously determined, supplemented by consistency checks of the
OPE as well as with other determinations of the bottom quark mass.

In this paper, we use the 1S mass $m_{1S}$ {as a short-distance mass. 
The 1S mass is originally defined as (perturbative part of)
one half of the mass of the 
bottomonium vector 1S state.
It is free of the $u=1/2$ renormalon.
See refs.~\cite{Hoang:1998hm,Hoang:1998ng} for the pioneering work to use the 1S mass for $|V_{cb}|$ determination. 
An advantage to use the 1S mass is that the input bottomonium
mass is precisely known from experiments. 
Since the bottom mass dependence appears as $(m_b^{\rm short})^5$ in eq.~\eqref{DecayWidth}
the error of $m_b^{\rm short}$ can induce a significant error to $|V_{cb}|$ determination.
We also consider it  advantageous that we do not need to introduce an artificial scale such as a factorization scale,
which generally induces certain instability of theoretical predictions.

We explain how to treat the OPE \eqref{DecayWidth} in our study.
In our study, we do not use the moments and
only use the inclusive semileptonic
decay width, for which the third-order perturbative series 
is available.
The central element for our theoretical prediction is the perturbative part $m_b^5 C_{\bar{Q}Q}^{\Gamma}$,
which we rewrite in terms of the 1S mass.
The perturbative series has the $u=1$ renormalon.
In this situation, since $\mu_{\pi}^2$ is an ambiguous non-perturbative effect due to the perturbative ambiguity of $\mathcal{O}(\LMS^2)$,
it is not clear how to appropriately include the $\mu_{\pi}^2$ effect in a theoretical prediction.
Also the $\mu_{\pi}^2$ effect is considered to be less than or similar to 
the perturbative error
since the achievable accuracy within perturbation theory, which is the $u=1$ renormalon uncertainty, is $\mathcal{O}(\LMS^2/m_b^2)$.
We then set $C_{\rm kin}^{\Gamma} \mu_{\pi}^2/m_b^2$ zero.
(We discuss this issue in more detail later.)
On the other hand, it is legitimate to include $\mu_G^2$,
since it is free of the uncertainty of $\mathcal{O}(\LMS^2)$ (as we explain below).
Furthermore, taking into account 
that its size is numerically large, we include $C_{\rm cm}^{\Gamma} \mu_G^2/m_b^2$ in our theoretical prediction.
In this analysis, we would like to clarify the achievable accuracy in the case one mainly uses the perturbative series alone
(without subtracting the $u=1$ renormalon).
Rephrasing this, we aim at clarifying the practical necessity to subtract the $u=1$ renormalon or to include 
the $\mu_{\pi}^2$ effect,
from the viewpoint of the current perturbative accuracy.

We determine the input value of the non-perturbative effect $\mu_G^2$ from 
the mass difference between the $B$ and the $B^*$ mesons.
The non-perturbative matrix elements included in eq.~\eqref{DecayWidth} 
appear also  in the OPE of the meson masses as
\be
m_B=m_b+\bar{\Lambda}+\frac{\mu_{\pi}^2}{2 m_b}-C_{\rm cm}^M \frac{\mu_G^2}{2 m_b} ,
\ee
\be
m_{B^*}=m_b+\bar{\Lambda}+\frac{\mu_{\pi}^2}{2 m_b}+C_{\rm cm}^M\frac{1}{3}\frac{\mu_G^2}{2 m_b} ,
\ee
up to higher order corrections in $1/m_b$.
As seen from the OPE, $\mu_G^2$ is the leading contribution to the difference
between $m_B$ and $m_{B^*}$.
It is thus free of the renormalon of $\mathcal{O}(\LMS^2)$  (since there are no quantities to cancel the renormalon if it existed).
We carry out a $\mu_G^2$ determination using the NNNLO perturbative result for $C_{\rm cm}^M$ \cite{Grozin:2007fh}.

The main novel points of the present paper are as follows.
We perform our $|V_{cb}|$ determinations not only using the $\Upsilon(1S)$ mass but also using the $\eta_b(1S)$ mass.
A determination using the $\eta_b(1S)$ mass is done for the first time.
We use the highest order perturbative series available today for $C_{\bar{Q}Q}^{\Gamma}$, i.e.,
the third-order perturbative series.
Our study makes it clear that there is a sizable difference of the determined $|V_{cb}|$
between the $\Upsilon(1S)$ mass scheme and the $\eta_b(1S)$ mass scheme.
This seriously limits the accuracy of $|V_{cb}|$ determination in the 1S mass scheme.
We discuss the cause of this undesired difference 
and conclude that it stems from the fact that the current perturbative calculation
for the mass splitting between $\Upsilon(1S)$ and $\eta_b(1S)$
tends to be much smaller than the experimental value.

The paper is organized as follows. In Sec.~\ref{sec:2},
we perform a determination of $\mu_G^2$ from the mass difference between the $B$ and the $B^*$ mesons.
In Sec.~\ref{sec:3} we perform $|V_{cb}|$ determinations in the $\Upsilon(1S)$ and the $\eta_b(1S)$ schemes,
and give our combined result.
In this analysis, the value of $\mu_G^2$ determined in Sec.~\ref{sec:2} is used.
In Sec.~\ref{sec:4}, we discuss the cause of the difference of $|V_{cb}|$ determinations between the two bottomonium 1S mass schemes.
Sec.~\ref{sec:5} is devoted to the conclusions.

\section{Determination of $\mu_G^2$}
\label{sec:2}
We determine the non-perturbative effect $\mu_G^2$ using the experimental values of $m_B$ and $m_{B^*}$.
We consider $m_{B^*}^2-m_B^2$, whose OPE is given by
\be
m_{B^*}^2-m_B^2=\frac{4}{3} C_{\rm cm}^M(\mu) \mu_G^2(\mu) , \label{massdiff}
\ee
where the input {{\it{bottom quark mass}} is unnecessary. 
The chromomagnetic operator $\bar{b}_v g_s \sigma_{\mu \nu} G^{\mu \nu} b_v$ has anomalous dimension
and $\mu_G^2(\mu)$ changes as $\mu$ varies.
We fix the renormalization scale to\fn{
$\overline{m}_b$ denotes the $\MSb$ mass whose renormalization scale is taken as $\mu=\overline{m}_b$.
$\overline{m}_c$ is understood in a parallel manner.}  $\mu=\overline{m}_b$
and determine $\mu_G^2(\mu=\overline{m}_b)$.
The Wilson coefficient $C_{\rm cm}^M$ was calculated to the $\mathcal{O}(\alpha_s^3)$ order \cite{Grozin:2007fh}.

We treat $C_{\rm cm}^M$ in the usual fixed-order perturbation theory.
We perform our $\mu_G^2$ determination noting that $C_{\rm cm}^M$ has a $u=1/2$ renormalon \cite{Grozin:1997ih}.\fn{
In the OPE of the mass difference $m_{B^*}-m_B$, the renormalon uncertainty of $C_{\rm cm}^M$ of $\mathcal{O}(\LMS/\overline{m}_b)$
is canceled against the uncertainty of the $\mathcal{O}(\LMS^3/m_b^2)$ contribution
in higher order terms.}
For the perturbative series of $C_{\rm cm}^M$,
we take the number of massless quarks as three rather than four.
This is because we expect that high-order perturbative coefficients are close to those of $C_{\rm cm}^{M (n_{\ell}=3)}$ rather than $C_{\rm cm}^{M (n_{\ell}=4)}$
when the internal charm mass effects are considered,
since the massive (charm) quark decouples from the renormalon asymptotic behavior \cite{Ball:1995ni}.
Simultaneously we consider the running coupling constant with three flavors.\fn{
We neglect the matching of the chromomagnetic operator to the three-flavor theory
since we expect that, for our purpose, i.e., $|V_{cb}|$ determination, its effect is negligible and the
three-loop relation is unknown. See refs.~\cite{Grozin:2006xm,Grozin:2000cm} for the known matching relations.
}
(Throughout this paper, we consider the running coupling constant with
the four-loop beta function.)

We show the result of $\mu_G^2$ in Fig.~\ref{fig:muGsq}
using the inputs $m_B=(5.27965+5.27934)/2$~GeV and $m_{B^*}=5.32470$~GeV.
We can neglect their small errors.
These numbers are the average values of the $B^0$ and $B^{\pm}$ mesons,
as we neglect the iso-spin breaking effects.
We also use $\overline{m}_b=4.18$~GeV and $\alpha_s^{(5)}(M_Z)=0.1179$, corresponding to $\LMS^{(n_{\ell}=3)}=0.332$~GeV \cite{Chetyrkin:1997sg} with
$\overline{m}_c=1.27$~GeV. 
These numbers, as well as the numbers in Sec.~\ref{sec:3}, are taken from 
the Particle Data Group
(PDG) values \cite{ParticleDataGroup:2020ssz}.
In the $\mu_G^2$ determination,
we also neglect the errors of these input values, 
since their effects are small compared to the dominant error 
coming from the error of $C_{\rm cm}^M$.
\begin{figure}[tbp]
\begin{center}
\includegraphics[width=10cm]{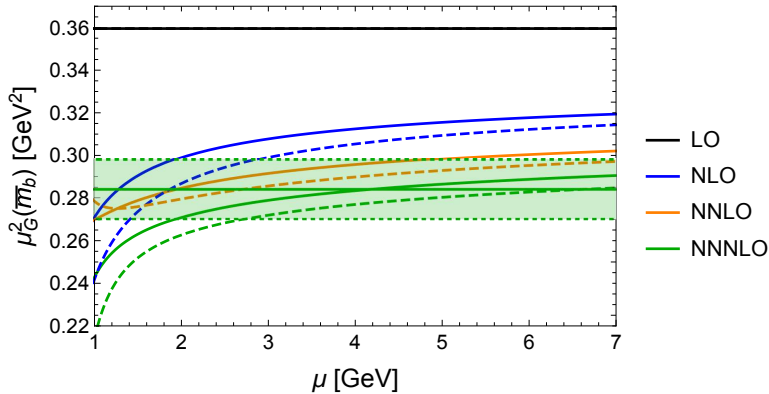}
\end{center}
\caption{Determination of $\mu_G^2$ using various order results for $C_{\rm cm}^M$.
The way to obtain the solid and the dashed lines is explained in the text.}
\label{fig:muGsq}
\end{figure}
In the figure, we examine the scale dependence of the ``RG invariant'' quantity $\mu_G^2(\overline{m}_b)$
(in the sense that the renormalization scale of the composite operator is already fixed)
by expressing the perturbative series for $C_{\rm cm}^M(\overline{m}_b)$ in terms
of $\alpha_s(\mu)$.
We show solid lines and dashed lines; the solid lines represent $\mu_G^2|_{{\rm N}^k{\rm LO}}=(m_{B^*}^2-m_B^2) \lt(\frac{4}{3} C_{\rm cm}|_{{\rm N}^k{\rm LO}} \rt)^{-1}$
while the dashed lines represent $\mu_G^2|_{{\rm N}^k{\rm LO}}=(m_{B^*}^2-m_B^2) \lt(\frac{4}{3} C_{\rm cm} \rt)^{-1}|_{{\rm N}^k{\rm LO}}$.
They are both possible ways to determine $\mu_G^2$ using the N$^k$LO Wilson coefficient,
and the difference between them provides us with a measure 
of the systematic uncertainty in the N$^k$LO analysis.
At NNNLO we do not find minimal sensitivity scales in a reasonable range of $\mu$.
We then fix our central value at a natural scale $\mu=\overline{m}_b$ using the solid line at NNNLO.
We estimate the perturbative error by changing the scale by the factor of $2$ or $1/2$.
This scale variation is done also for the NNNLO dashed line. 
The maximal variation is taken as our error.
We obtain
\be
\mu_G^2(\overline{m}_b)=0.284 \pm 0.014~{\rm GeV}^2 .
\ee
The central value and the error are shown by the green band in the figure.
The difference from the previous order (NNLO) is of the same order as the assigned error.

As seen from the figure, the behavior of $\mu_G^2$ is not stable against inclusion of higher orders.
We infer that this is because of the $u=1/2$ renormalon of $C_{\rm cm}^M$.
Our error in the NNNLO analysis is order 5~\%, 
and this is the same order of magnitude as the naively expected renormalon uncertainty of $C_{\rm cm}^M$, that is, $\LMS/\overline{m}_b \sim 8~\%$.
In fact this error size is sufficiently small for the determination of $|V_{cb}|$.
For a more precise determination of $\mu_G^2$, 
subtraction of the $u=1/2$ renormalon from $C_{\rm cm}^M$ is required.

There are previous determinations of $\mu_G^2$; 
see, e.g., refs.~\cite{Alberti:2014yda,FermilabLattice:2018est,Gambino:2017dfa,Fael:2020tow} for recent determinations.
In ref.~\cite{Gambino:2017dfa}, $\mu_G^2(\mu=\tilde{m}_b=4.605~{\rm GeV})=0.253(21)_{\text{stat+syst}}(13)_{\text{conv}}$~GeV$^2$
is obtained from the mass difference between the $B$ mesons, and in ref.~\cite{Alberti:2014yda},
$\mu_G^2(\mu=4.6~{\rm GeV})=0.333(61)$~GeV$^2$ is obtained from the semileptonic $B$ decay.
To compare our result with these values, we give 
\be
\mu_G^2(\mu=4.6~{\rm GeV})
=\exp \lt[\int_{\alpha_s(\overline{m}_b)}^{\alpha_s(4.6~{\rm GeV})} dx \, \frac{\gamma_{\rm cm}(x)}{2 \beta(x)} \rt] \mu_G^2(\overline{m}_b)
=0.287 \pm 0.014~{\text{GeV}}^2 ,
\ee
where $\gamma_{\rm cm}(x)$ is the anomalous dimension of $\mu_G^2$ given in ref.~\cite{Grozin:2007fh}.
(We define $\beta(\alpha_s)=\mu^2 d \alpha_s/d \mu^2$.)

\section{Determination of $|V_{cb}|$}
\label{sec:3}
In this analysis, we use the 1S mass for the bottom quark mass and
the $\MSb$ mass for the charm quark mass. 
We define $m_{\rm 1S}$ by one half of the mass of a bottomonium 1S state
and $m_{\rm 1S}$ means $m_{\Upsilon(1S)}$ or $m_{\eta_b(1S)}$ here. 
We use the relation between
the pole and the 1S masses, which is given in the so-called $\epsilon$ expansion \cite{Hoang:1998hm,Hoang:1998ng,Hoang:2000fm},
\be
\frac{m_{\rm 1S}}{m_b}=1+\sum_{n=1}^3 \epsilon^n e_n(\alpha_s, \overline{m}_c/m_{\rm 1S}) ,
\ee
instead of the usual $\alpha_s$ expansion. 
The $\epsilon$ expansion is a convenient framework 
which can treat the $u=1/2$ renormalon properly, whereas
the order counting in $\alpha_s$ is not appropriate in the situation that the typical scale or the Bohr radius of
the bottomonium depends on $\alpha_s$ as $ ( \frac{1}{2} C_F \alpha_s m_b)^{-1}$.  
Each coefficient of $\epsilon^{n}$ depends on $\alpha_s$.
The parameter $\epsilon$ is finally set to one.
For the mass relation, we basically use the results in ref.~\cite{Hoang:2000fm}
but we consider the internal charm mass effects beyond the linear approximation in $\overline{m}_c/m_{\rm 1S}$
using the results in ref.~\cite{Fael:2020bgs}.
For the charm quark, we use the relation between the pole and the $\MSb$ masses 
up to the $\mathcal{O}(\alpha_s^3)$ order 
\cite{Tarrach:1980up,Gray:1990yh,Chetyrkin:1999ys,Chetyrkin:1999qi,Melnikov:2000qh,Melnikov:2000zc,Marquard:2015qpa,Marquard:2016dcn}.
We include non-decoupling effects of the bottom quark using the results in ref.~\cite{Fael:2020bgs},
although the effects turned out to be small. (The numerical impact of the inclusion of the effects is explicitly given in App.~A of ref.~\cite{Hayashi:2021vdq}.)

We make a remark on the $u=1/2$ renormalon cancellation in the product $m_b^5 C_{\bar{Q}Q}^{\Gamma}$.
Since the Wilson coefficient $C_{\bar{Q}Q}^{\Gamma}$ is calculated by treating the charm quark as massive, 
its renormalon uncertainty is actually that of three-flavor QCD. 
To cancel the renormalon in the product $m_b^5 C_{\bar{Q}Q}^{\Gamma}$, the renormalon uncertainty of $m_b$ should also be that of three-flavor QCD. 
We therefore need to consider internal charm mass effects on the mass relation for consistency \cite{Ball:1995ni,Hayashi:2021vdq}.
Related to this, we consider the three-flavor running coupling constant 
as it is natural in this situation \cite{Ayala:2014yxa,Hayashi:2021vdq}. 

Following the above procedure, we can express the perturbative part of the decay width as
\be
m_b^5 C_{\bar{Q}Q}^{\Gamma} (m_c/m_b, \alpha_s)=m_{\rm 1S}^5 \sum_{n= 0}^3 \epsilon^n f_n(\overline{m}_c/m_{\rm 1S},\alpha_s) .
\ee
The perturbative series $\sum_{n= 0}^3 \epsilon^n f_n(\overline{m}_c/m_{\rm 1S},\alpha_s)$
is free of the $u=1/2$ renormalon but has the $u=1$ renormalon.
For the $\mu_G^2$ part, we use the LO result of $C_{\rm cm}^{\Gamma}$,
\be
C_{\rm cm}^{\Gamma}|_{\rm LO}=-\frac{1}{2} (3-8\rho+24 \rho^2-24 \rho^3+5 \rho^4+12 \rho^2 \log{\rho})
\ee  
with $\rho=\overline{m}_c^2/m_{\rm 1S}^2$.
Then we determine $|V_{cb}|$ by 
\be
|V_{cb}|=\Gamma_{\exp}^{1/2} \lt(\frac{G_F^2 A_{\rm EW}}{192 \pi^3}  \rt)^{-1/2} \sqrt{\frac{1}{m_{\rm 1S}^5}}
\sqrt{\frac{1}{\sum_{n=0}^3 \epsilon^n f_n(\overline{m}_c/m_{\rm 1S},\alpha_s)+C_{\rm cm}^{\Gamma}|_{\rm LO} \frac{\mu_G^2}{m_{\rm 1S}^2} }} .
\ee
The results are shown by the solid lines in Fig.~\ref{fig:Vcb} as 
the truncation order in $\epsilon$ is changed.
We also consider the short-distance expansion of the right-hand side;
we first expand in $C_{\rm cm}^{\Gamma}|_{\rm LO} \frac{\mu_G^2}{m_{\rm 1S}^2}$ to obtain
$\sqrt{\frac{1}{\sum_{n=0}^3 \epsilon^n f_n(\overline{m}_c/m_{\rm 1S},\alpha_s)}}-\frac{1}{2} \lt(\frac{1}{f_0(\overline{m}_c/m_{\rm 1S})} \rt)^{3/2} C_{\rm cm}^{\Gamma}|_{\rm LO} \frac{\mu_G^2}{m_{\rm 1S}^2} $
and then expand $\sqrt{\frac{1}{\sum_{n=0}^3 \epsilon^n f_n(\overline{m}_c/m_{\rm 1S},\alpha_s)}}$ in $\epsilon$ to a given order.
The results obtained in this way are shown by the dashed lines.
\begin{figure}[tb]
\begin{minipage}{0.5\hsize}
\begin{center}
\includegraphics[width=8cm]{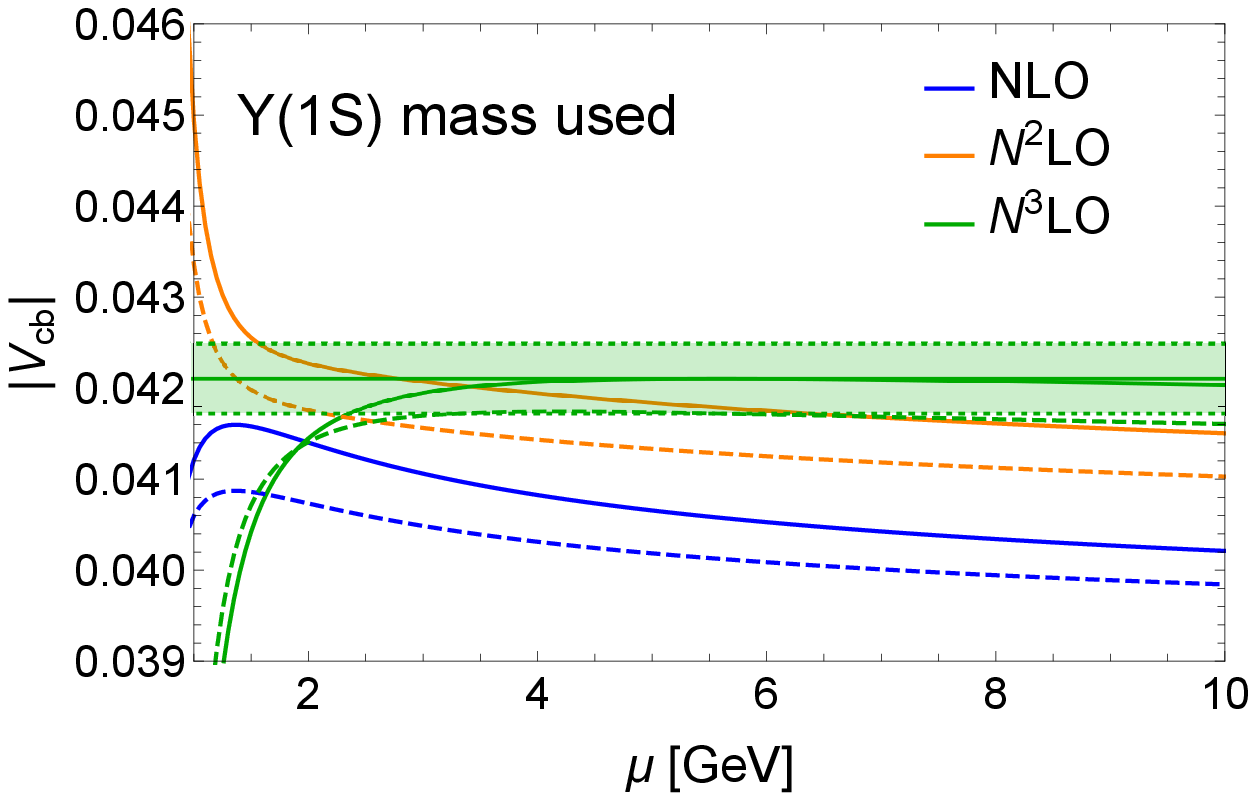}
\end{center}
\end{minipage}
\begin{minipage}{0.5\hsize}
\begin{center}
\includegraphics[width=8cm]{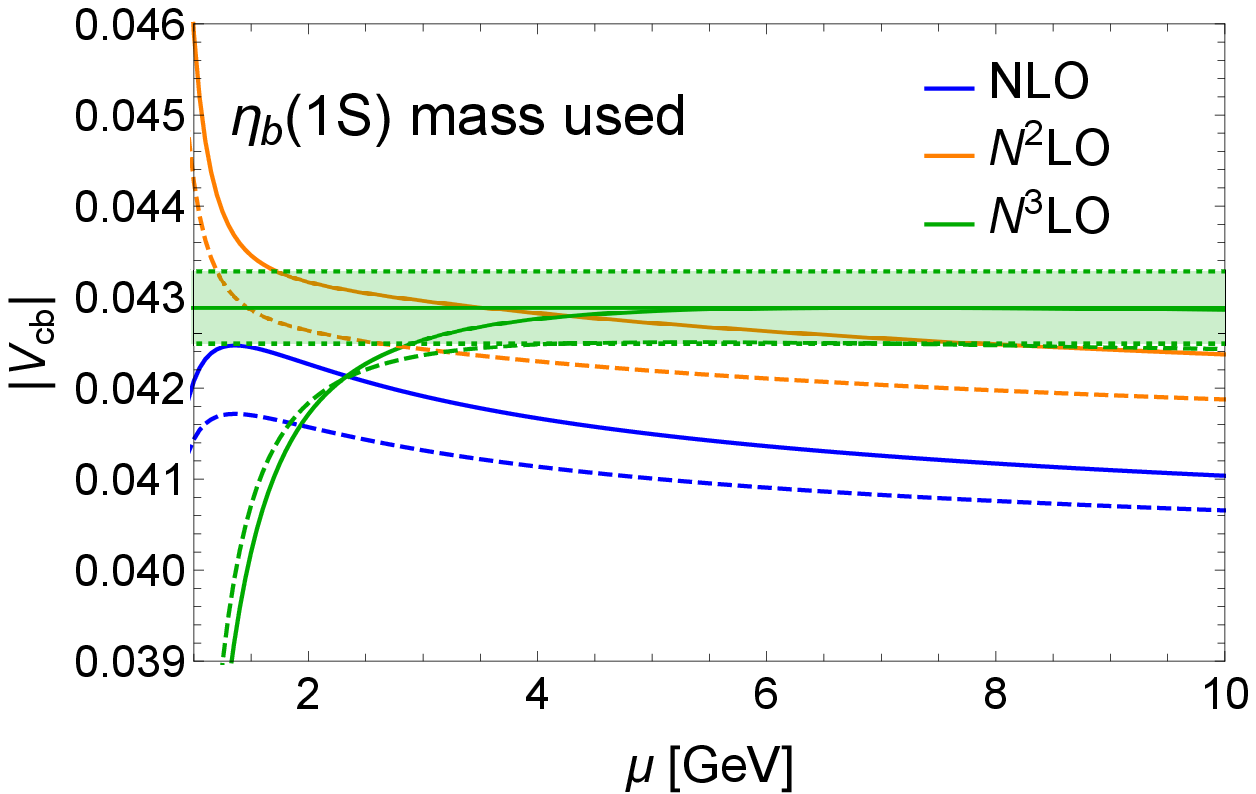}
\end{center}
\end{minipage}
\caption{Determination of $|V_{cb}|$ using the $\Upsilon(1S)$ mass (left) and $\eta_b(1S)$ mass (right).
We examine scale dependence and stability against including higher orders. The ways to obtain 
the solid and the dashed lines are explained in the text. A green band shows the central value and the perturbative error 
in each determination.}
\label{fig:Vcb}
\end{figure}

We list the values of the main input parameters and their errors in Table~\ref{tab:inputs}. 
In Fig.~\ref{fig:Vcb}, we use the central values.
We also use $G_F=1.1663787 \times 10^{-5}$~GeV$^{-2}$ and $A_{\rm EW}=1.014$ \cite{Benson:2003kp,Sirlin:1974ni}.
We neglect the errors of these values.
$\Gamma_{\rm exp}$ is given by $\Gamma_{\rm exp}={\rm Br}/ \tau_B$, where $\tau_B$ is the lifetime of the $B$ mesons
and ${\rm Br}$ denotes the branching ratio to the semileptonic decay.\fn{
In this analysis, we neglect the iso-spin breaking and assume that the semileptonic decay width of $B^0$ and that of $B^{\pm}$ is the same.
Then we can obtain the semileptonic decay width by ${\rm Br} /\tau_B$ 
where ${\rm Br}$ is the semileptonic branching ratio obtained for the admixture of $B^0/B^{\pm}$ (which is given in Table~\ref{tab:inputs})
and $\tau_B=(\tau_{B^{\pm}}+\tau_{B^0})/2+\frac{1}{2}(f^{+-}-f^{00}) (\tau_{B^{\pm}}-\tau_{B^0})$,
as clearly explained in Sec.~III~A of ref.~\cite{Bauer:2004ve}.
Here $f^{+-}$ and $f^{00}$ are the fractions of the $B^+ B^-$ production 
and the $B^0 \bar{B}^0$ production from the $\Upsilon(4S)$ decay, respectively.
We give the value $\tau_B$ in Table~\ref{tab:inputs} neglecting the second term since it is smaller than the error of $\tau_{B^{\pm}}$ or $\tau_{B^0}$.
See ref.~\cite{HFLAV:2019otj} for the ratio $f^{+-}/f^{00}$.
}
\begin{table}[t]
\begin{center}
\begin{tabular}{c|c}
\hline 
Parameter &   \\
\hline \hline
$\alpha_s(M_Z) $             &            $0.1179\pm 0.0010$ \\ 
$\overline{m}_c$               &            $1.27 \pm 0.02$~GeV \\ 
$\mu_G^2$                       &            $0.284 \pm 0.014$~GeV$^2$ \\ 
$m_{\Upsilon(1S)}$          &            $(9.46030 \pm 0.00026)/2$~GeV  \\ 
$m_{\eta_b(1S)}$             &            $(9.3987 \pm 0.0020)/2$~GeV \\ 
Br                 &            $(10.65 \pm 0.16) \times 10^{-2}$ \\
$\tau_B$                               &            $[(1.519\pm0.004)+(1.638 \pm 0.004)]/2 \times 10^{-12}$~sec \\
\hline
\end{tabular}
\end{center}
\caption{Values of the input parameters.}
\label{tab:inputs}
\end{table}

We first explain the case we use $m_{\rm 1S}=m_{\Upsilon(1S)}$.
We see in Fig.~\ref{fig:Vcb} that the scale dependence gets milder
as the perturbation order is raised.
We mainly focus on the solid line at NNNLO. We find the minimal sensitivity scale at $\mu_0 \simeq 5.6$~GeV,
where $\alpha_s(\mu_0) \simeq 0.1950$ and the perturbative series behaves as 
$\sum_{n= 0}^3 \epsilon^n f_n(\overline{m}_c/m_{\rm 1S},\alpha_s)=0.5903-0.0836 \epsilon-0.0281 \epsilon^2-0.0070 \epsilon^3$.
We determine our central value in this analysis at this scale and obtain $|V_{cb}|=42.1 \times 10^{-3}$.

To estimate the perturbative error, we perform the following analysis.
(i) We take the difference from the NNLO solid line at the minimal sensitivity scale $\mu_0 \simeq 5.6$~GeV.
(ii) We take the difference from the NNNLO dashed line at the minimal sensitivity scale $\mu_0 \simeq 5.6$~GeV.
(iii) We change the renormalization scale within $ [\mu_0/2, 2\mu_0]$ and examine variations of the NNNLO solid line.
We determine the perturbative error by the maximal deviation among them.
We obtain $\pm 0.4 \times 10^{-3}$ from (ii) as the perturbative error. 
The central value with the perturbative error is shown by the green band.
One can check validity of the error from the lines in the figure.
We find that the NNNLO result indeed has a smaller error
compared to the NNLO result if we use the same estimation method.

We vary the input parameters or the experimental values within their errors independently to examine individual errors.
In estimating the error concerning $\tau_B$, we take the possible maximum deviation neglecting the correlation between their errors.

We also roughly estimate the impact of the NLO correction to $C_{\rm cm}^{\Gamma}$.
We use the numerical result given in Sec.~4 of Ref.~\cite{Alberti:2013kxa}.
Since the given number is obtained in a different mass scheme, this analysis just 
indicates a rough size of the effect.

As the final error source, we discuss the possible impact of the $u=1$ renormalon
or that of $\mu_{\pi}^2$, which is deeply related to each other.
In our analysis, the $u=1$ renormalon is not subtracted.
If the renormalon uncertainty is non-zero, it should be approximately less than or equal to the perturbative error,
since the renormalon uncertainty is the minimal error reached within 
the fixed-order perturbation theory.
In this sense, we have already estimated its impact. 
However, the estimate of the $u=1$ renormalon effect is not sufficient 
to estimate the $\mu_{\pi}^2$ effect if the unambiguous part of $\mu_{\pi}^2$, which remains after the $u=1$ renormalon cancellation
and contributes to the decay width, is larger than the $u=1$ renormalon uncertainty.
This possibility is relevant particularly in this case because the $u=1$ renormalon is considered to be small \cite{Beneke:1994sw,Ayala:2019hkn,Beneke:2021lkq}.
To estimate the effect of $\mu_{\pi}^2$,
we refer to our previous estimate of the unambiguous part of $\mu_{\pi}^2$, given by  $\mu_{\pi}^2=-0.12 \pm 0.23$~GeV$^2$ \cite{Hayashi:2021vdq}.
Varying $\mu_{\pi}^2$ within this range and using the Wilson coefficient $C_{\rm kin}^{\Gamma}$ at LO,
we estimate the impact of the neglected $\mu_{\pi}^2$.

Our result in the case using $m_{\Upsilon(1S)}$ reads
\begin{align}
|V_{cb}|
&=42.1 (4)_{\rm pert} (1)_{\alpha_s} (4)_{\overline{m}_c} (0)_{\mu_G^2} (0)_{m_{\Upsilon(1S)}} (3)_{\rm Br} (1)_{\tau_B} (1)_{\text{h.o.$C_{\rm cm}^{\Gamma}$}} (2)_{\rm \mu_{\pi}^2} \times 10^{-3} \non
&=42.1 (7) \times 10^{-3} , \label{Upsilonres}
\end{align}
where each systematic error is shown and they are combined in quadrature in the final equality. 
The error of the input bottomonium mass is indeed negligible. 
This is an advantage of the 1S mass scheme.
However, we note that so far
we have considered only the perturbative relation between the pole 
(or $\MSb$) mass
and the 1S mass, and neglected non-perturbative corrections.
We discuss this issue in Sec.~\ref{sec:4}.

The error caused by neglecting $\mu_{\pi}^2$ is about half of the perturbative error.
This error can be removed only if one subtracts the $u=1$ renormalon
from the perturbative series and includes the unambiguous part of $\mu_{\pi}^2$ in the OPE.
This indicates that the order of the perturbative calculation is becoming so high
that it is necessary to include the non-perturbative effect in an appropriate way. 
We also examine the significance to include $\mu_G^2$. Neglecting the $\mu_G^2$ term by
setting $\mu_G^2=0$, we observe the shift $|V_{cb}|=42.1 \times 10^{-3} \to 41.4 \times 10^{-3}$.
The variation is larger than the perturbative error and shows clear necessity to include $\mu_G^2$.
In fact, this non-perturbative effect contributes as $C_{\rm cm}^{\Gamma} \mu_G^2/m_{\Upsilon(1S)}^2 \simeq -0.0151$
and its magnitude is larger than the highest perturbative correction in the series $0.5903-0.0836 \epsilon-0.0281 \epsilon^2-0.0070 \epsilon^3$.

We now perform a parallel analysis using the $\eta_b(1S)$ mass.
In this analysis, we use the results of ref.~\cite{Kiyo:2014uca} as well, for the relation between the pole and the 1S masses.
The minimal sensitivity scale is found at $\mu_0 \simeq 7.1$~GeV, where 
$\alpha_s(\mu_0)=0.1816$ and the perturbative series behaves as $0.5863-0.0797 \epsilon-0.0294 \epsilon^2-0.0073 \epsilon^3$.
(See Fig.~\ref{fig:Vcb}.)
Our result using the $\eta_b(1S)$ mass reads
\begin{align}
|V_{cb}|
&=42.9 (4)_{\rm pert} (4)_{\overline{m}_c} (1)_{\alpha_s} (0)_{\mu_G^2} (0)_{m_{\eta_b(1S)}} (3)_{\rm Br} (1)_{\tau_B} (0)_{\text{h.o.$C_{\rm cm}^{\Gamma}$}}  (2)_{\rm \mu_{\pi}^2} \times 10^{-3} \non
&=42.9 (7) \times 10^{-3} . \label{Etabres}
\end{align}
We find a shift of the central values in eqs.~\eqref{Upsilonres} and \eqref{Etabres} by $0.8 \times 10^{-3}$.

We give our final central value by the average of the two central values,
and give the systematic error concerning the ones explained above by the average of the total errors.
In addition, since we find a sizable difference of the central values between the two cases,
we take the difference as the systematic error concerning the 
bottomonium spin dependence.
Our final result is given by
\begin{align}
|V_{cb}|
&=42.5 (7)_{\text{averaged sys.~error}} (8)_{\text{spin diff.}} \times 10^{-3} \non
&=0.0425 (11) .
\label{finalresult}
\end{align}

\section{Difference between the $\Upsilon(1S)$ and $\eta_b(1S)$ schemes}
\label{sec:4}

In this section, we argue that the large difference of the central values of $|V_{cb}|$ between the two 1S mass schemes
stems from the fact
that the current perturbative result
for the mass splitting between $\Upsilon(1S)$ and $\eta_b(1S)$
tends to be much smaller than the observed  value.
In fact, the perturbative result reads
\be
\frac{m^{\rm NNNLO}_{\Upsilon(1S)}}{m^{\rm NNNLO}_{\eta_b(1S)}}= 0.994 \frac{m^{\rm exp}_{\Upsilon(1S)}}{m_{\eta_b(1S)}^{\rm exp}}   . \label{deviation}
\ee
Here we used 
\be
\frac{m_{\Upsilon(1S)}}{\overline{m}_b} \bigg|_{\rm NNNLO}=1+0.0810 \epsilon+0.0303 \epsilon^2+0.0116 \epsilon^3=1.1229, \label{UpsilonMass}
\ee
\be
\frac{m_{\eta_b(1S)}}{\overline{m}_b}\bigg|_{\rm NNNLO}=1+0.0810 \epsilon +0.0303 \epsilon^2+0.0105 \epsilon^3=1.1218 , \label{EtabMass}
\ee
taking $\mu=\overline{m}_b=4.18$~GeV and $\overline{m}_c/\overline{m}_b=1.27/4.18$ as a reference point.
The same tendency was indeed observed in ref.~\cite{Kiyo:2015ufa}.
This deviation causes the difference of the central values of $|V_{cb}|$.
To be explicit we rewrite the leading perturbative contribution to the decay width 
\be
\overline{m}_b^5 C^{\Gamma}_{\bar{Q}Q}|_{\rm LO}
=\overline{m}_b^5 (1-8 \bar{\rho}+8 \bar{\rho}^3- \bar{\rho}^4-12 \bar{\rho}^2 \log{\bar{\rho}}) \quad{} \text{with $\bar{\rho}=\overline{m}_c^2/\overline{m}_b^2$}
\ee
converting $\overline{m}_b$ into either $m_{\Upsilon(1S)}$ or $m_{\eta_b(1S)}$ by using eq.~\eqref{UpsilonMass} or eq.~\eqref{EtabMass}.
We obtain
\be
\frac{\overline{m}_b^5 C^{\Gamma}_{\bar{Q}Q}|_{{\rm LO}} [\Upsilon(1S)~\text{scheme}] }{\overline{m}_b^5 C^{\Gamma}_{\bar{Q}Q}|_{\rm LO} [\eta_b(1S)~\text{scheme}]}
=1.034 , \label{DeviationLOGamma}
\ee
where we used the experimental values for $m_{\Upsilon(1S)}$ and $m_{\eta_b(1S)}$.
From eq.~\eqref{DeviationLOGamma}, we expect that the value of $|V_{cb}|$ gets larger in the $\eta_b(1S)$ scheme
than in the $\Upsilon(1S)$ scheme by $\sim 3.4/2~\% \sim 1.7$~\%, which corresponds to a shift of $+0.7 \times 10^{-3}$.
This is indeed close to the obtained shift of $+0.8 \times 10^{-3}$.\footnote{
A similar estimate is obtained if we use the pole mass instead of $\overline{m}_b$
(by subtracting the renormalons of the pole mass using eq.~(5.33) of ref.~\cite{Hayashi:2021vdq}.)
} 
Therefore we conclude that the large shift (with a definite sign) is caused by the scheme dependence or the specific tendency 
which is represented by eq.~\eqref{deviation}.

In view of the above analysis, 
let us discuss the current status of the 1S mass schemes.
Our analysis in Sec.~\ref{sec:3} is based on a relation between two physical observables,
i.e., $\Gamma(B \to X_c \ell \nu)$ and the $\Upsilon(1S)$ mass, or $\Gamma(B \to X_c \ell \nu)$ and the $\eta_b(1S)$ mass.
Prior to this analysis, there have been examinations of the
relations between $\overline{m}_b$ and the masses of $\Upsilon(1S)$
and  $\eta_b(1S)$ \cite{Kiyo:2015ufa,Peset:2018ria}, where it has been estimated that the
non-perturbative corrections to these relations are
smaller or comparable to the current perturbative error.\footnote{
The leading non-perturbative contribution to the bottomonium spectrum
is given by the non-local gluon condensate which represents contributions
from the ultrasoft scale and smaller scales \cite{Brambilla:1999qa}.
This contribution is spin independent and
is estimated to be smaller than or comparable to
the current perturbative error to each energy level. 
This feature is further confirmed in the analysis of the static potential
\cite{Takaura:2018vcy}.
} 
Based on this observation, in Sec.~\ref{sec:3} we have not included 
non-perturbative corrections to the bottomonium 1S masses,
while we have included a non-perturbative correction ($\mu_G^2$) to the $B$ semileptonic width.
The central value of the current (fixed-order) perturbative calculation for the 
bottomonium 1S mass splitting $\Delta M_{b\bar{b}({\rm 1S})}$ 
is known to be considerably smaller than the corresponding
experimental value.
There exist indications that higher-order perturbative corrections
may resolve this discrepancy \cite{Recksiegel:2003fm,Kniehl:2003ap},
but currently it remains an open question whether large non-perturbative
corrections exist in this splitting.
It requires calculation of the higher-order perturbative corrections
to $\Delta M_{b\bar{b}({\rm 1S})}$ in order to answer this question unambiguously.

Hence, the conclusion which can be drawn at the current status is either
(or combination) of the following:
(i) non-perturbative corrections to $\Delta M_{b\bar{b}({\rm 1S})}$
resolves the discrepancy of $|V_{cb}|$ in the two 1S schemes;
(ii) higher-order perturbative corrections to $\Delta M_{b\bar{b}({\rm 1S})}$
resolves the discrepancy of $|V_{cb}|$.
In the case (ii), the perturbative error in Sec.~\ref{sec:3}
must be underestimated by a factor of two.
Although the estimate method we used there meets today's standard,
it is possible that such an underestimate could occur accidentally.
Taking all these into account, we stress that irrespective of the case (i) or (ii), 
the final error size given in eq.~\eqref{finalresult}
should be  reasonable, since we have included the discrepancy in
the error.

\section{Conclusions}
\label{sec:5}
We performed a $|V_{cb}|$ determination using the recently calculated third-order
perturbative series and the 1S mass as a short-distance mass.
For the first time we used $m_{\eta_b(1S)}$ in addition to $m_{\Upsilon(1S)}$, 
while the latter has been exclusively used in the literature.
We observed reasonable convergences of the perturbative series in the two schemes
and the perturbative errors get smaller than those expected at the previous order.
In our study, we found that there is a sizable difference of $|V_{cb}|$ between the two determinations 
using the $\Upsilon(1S)$ mass and the $\eta_b(1S)$ mass. 
This gives the dominant error to our determination.
Our result reads $|V_{cb}|=0.0425(11)$, which is consistent with the PDG value determined from the inclusive decays 
$|V_{cb}|_{\rm PDG}=0.0422(8)$ and has a slightly larger error.

We find that the sizable difference in the two determinations originates from the fact that
the current perturbative prediction tends to
fall short in explaining 
the size of the mass splitting between $\Upsilon(1S)$ and $\eta_b(1S)$.
It requires calculation of the higher-order perturbative corrections
to this splitting in order to clarify whether higher-order perturbative
corrections or non-perturbative corrections are responsible to resolve this
discrepancy.
We have thus revealed that the 1S mass scheme has an intrinsic
uncertainty previously not attended to in the determination of $|V_{cb}|$.
Furthermore,
it is notable that the value of $|V_{cb}|$ which we obtained is consistent with
the PDG value extracted from the inclusive $B$ decays, while
it is in tension with that from the exclusive decays [$0.0395(9)$],
even after we take into account this uncertainty from the
bottomonium 1S mass splitting.

In our analyses, we paid attention to the effective number of flavors
to properly treat $\mathcal{O}(\LMS)$ renormalon uncertainties.
Since the charm quark is treated as a massive quark in the calculation of $C_{\bar{Q}Q}^{\Gamma}$,
we consistently consider internal charm mass effects in rewriting $m_b$ in terms of the short-distance mass $m_{\rm 1S}$.
In this way, the $\mathcal{O}(\LMS)$ renormalon uncertainties of $C_{\bar{Q}Q}^{\Gamma}$ and $m_b$ are both those
of three-flavor QCD, and the renormalon cancellation in the decay width accurately holds.
In accordance with this, we used the three-flavor running coupling constant.

We also estimated the impact of the $\mathcal{O}(\LMS^2)$ non-perturbative effects, $\mu_{\pi}^2$ and $\mu_G^2$, 
on determination of $|V_{cb}|$.
Our theoretical prediction is free from the $\mathcal{O}(\LMS)$ renormalon but not from the $\mathcal{O}(\LMS^2)$ renormalon.
Since $\mu_G^2$ is known  to be free of the $\mathcal{O}(\LMS^2)$ renormalon, 
we included only $\mu_G^2$ in our theoretical prediction.
We estimated $\mu_G^2$ from the mass difference between the $B$ and the $B^*$ mesons
using the NNNLO result for the Wilson coefficient $C_{\rm cm}^M$ in eq.~\eqref{massdiff}.
We found that this non-perturbative effect has a more significant contribution to the decay width 
than the third-order perturbative correction while it is less significant than the second-order correction.
We also estimated the possible impact of the neglected $\mu_{\pi}^2$
and found that it can be similar size to the current perturbative error.
These estimates clarify that proper inclusion of these non-perturbative effects has become relevant
in the present situation where the third order of the perturbative series is available.

It would be possible to remove the $u=1$ renormalon from the perturbative series using 
renormalon subtraction methods.
We intend to perform a $|V_{cb}|$ determination by 
subtracting the $\mathcal{O}(\LMS^2)$ renormalon
using our method \cite{Takaura:2020byt,Hayashi:2021vdq} in the near future.

\section*{Acknowledgments}
The work of Y.H. was supported by Grant-in-Aid for JSPS Fellows (No. 21J10226) from MEXT
and he also acknowledges support from GP-PU at Tohoku University.
This work was also supported by Grants-in-Aid for Scientific Research numbers 
JP20K03923 (Y.S.) and JP19K14711 (H.T.).

\bibliographystyle{utphys}
\bibliography{BibQCD}

\end{document}